\documentclass[useAMS,usenatbib]{mn2e}
\usepackage{epsfig}
\usepackage{aas_macros}     
\usepackage{amsfonts}
\title[Redshift space distortions
]{Modelling redshift space distortions in hierarchical cosmologies
}
\author[E.~Jennings, C.~M.~Baugh, S.~Pascoli]
{E. Jennings$^{1,2}$\thanks{E-mail: elise.jennings@durham.ac.uk}, C. M. Baugh$^{1}$, S. Pascoli$^{2}$\\
$^{1}$ Institute for Computational Cosmology, Department of Physics,  Durham University, South Road, Durham, DH1 3LE, U.K.\\
$^{2}$ Institute for Particle Physics Phenomenology, Department of Physics, Durham University, South Road, Durham, DH1 3LE, U.K.\\
}
\begin{document}

\date{}

\pagerange{\pageref{firstpage}--\pageref{lastpage}} \pubyear{2010}

\maketitle

\label{firstpage}

\begin{abstract}
The anisotropy of clustering in redshift space provides a direct measure of the growth rate of large scale structure in the Universe.
Future galaxy redshift surveys will make high precision measurements of these distortions, and will potentially allow us to distinguish between different 
scenarios
for the accelerating expansion of the Universe. Accurate predictions are needed in order to distinguish between competing
cosmological models.
 We study the distortions 
in the redshift space power spectrum in $\Lambda$CDM and quintessence dark energy models, using large volume N-body simulations, and
test predictions for the form of the  
redshift space distortions.
We find that the linear perturbation theory prediction by Kaiser (1987)  is a poor fit to the measured distortions, even 
on  surprisingly large scales $k \ge 0.05 h$Mpc$^{-1}$.
 An improved model for the redshift space power spectrum, including the 
non-linear velocity divergence power spectrum, is presented and
 agrees with the power spectra measured from the simulations
up to $k \sim 0.2 h$Mpc$^{-1}$. 
We have found a density-velocity relation which is cosmology independent and which relates the 
non-linear velocity divergence spectrum to the non-linear matter power spectrum. 
We provide a formula which generates the non-linear velocity divergence 
$P(k)$ at any redshift, using only the non-linear matter power spectrum and the linear growth factor at the desired redshift. 
This formula is accurate to better than 
$5\%$ on scales $k<0.2 h $Mpc$^{-1}$ for all the 
cosmological models discussed in this paper. Our results will extend the statistical power of future galaxy surveys.

\end{abstract}


\section{Introduction}

The rate at which cosmic structures grow is set by a competition between gravitational instability and the rate of expansion of the Universe. 
The growth of structure can be measured by analysing the distortions in the galaxy clustering pattern, when viewed in redshift space (i.e. when a galaxy's
redshift is used to infer its radial position).
Proof of concept of this approach came recently from 
%
\citet{Guzzo:2008ac} who used spectroscopic data for 10,000 galaxies from the VIMOS-VLT Deep Survey \citep{LeFevre:2004hv} 
to measure the growth rate of structure
at redshift $z=0.77$ to an accuracy of $\sim 40\%$ \citep[see also][]{2001Natur.410..169P}. 
To distinguish between competing explanations for the accelerating expansion of the Universe, we need to measure the growth of structure to an accuracy of a few 
percent over a wide redshift interval. 
The next generation of galaxy redshift surveys, such as ESA's Euclid mission \citep{2009ExA....23...39C}, will be able to achieve this precision.  
These redshift space distortions are commonly modelled using a linear perturbation theory expression.
We test the validity of this approximation using large volume N-body simulations
 to model the redshift space distortions in $\Lambda$CDM and  quintessence dark energy models, to see if it works at the level required to take advantage
of the information in forthcoming surveys. 
The large
volume of our simulations means that we are able to find the limits of  perturbation theory models.
We can also study the impact of non-linearities on large scales 
in cosmologies with different expansion histories from $\Lambda$CDM, such as quintessence dark energy.

One explanation of the accelerating expansion of the Universe is that a negative pressure dark energy component
makes up approximately 70\% of the present density of the Universe
\citep{Komatsu:2008hk, 2009MNRAS.400.1643S}.
Examples of dark energy models include the cosmological constant and a dynamical scalar field such as quintessence \citep[ see e.g.][for a review]{Copeland:2006wr}.
Other possible solutions require modifications to  general relativity and
 include extensions to the Einstein-Hilbert action,
such as $f(R)$ theories or braneworld cosmologies \citep[see e.g.][]{Dvali:2000hr,Oyaizu:2008sr}.

The expansion history of the Universe is described by the scale factor, $a(t)$.
Dark energy and modified gravity models can produce similar expansion histories for the Universe, which can be derived
from the Hubble parameter measured, for example, using type Ia supernovae.
As both dark energy and modified gravity models can be described using an effective equation of state which specifies 
the expansion history, it is not possible to distinguish between
these two possibilities using measurements of the expansion history alone.

The growth rate is a measure of how rapidly structures are forming in the Universe.
Dark energy or modified gravity models predict  different growth rates for the large scale structure of the Universe, which can be measured 
using redshift space distortions of clustering. 
As noted by \citet{Linder:2005in}, in the case of general relativity, the second order differential equation for the growth of density perturbations depends only on the
expansion history through the Hubble parameter, $H(a)$, or the equation of state, $w(a)$. This is not the case for modified gravity theories.
By comparing the cosmic expansion history with the growth of structure, it is possible to distinguish the physical origin of
the accelerating expansion
of the Universe as being due either to dark energy or modified gravity \citep*{PhysRevD.69.124015,Linder:2005in}.
If there is no discrepancy between the observed growth rate and the theoretical
prediction assuming general relativity, this implies that a dark energy component alone can explain the accelerated expansion.

Galaxy redshift surveys allow us to study the 3D spatial distribution of galaxies and clusters. In a homogeneous universe,
redshift measurements would probe only the Hubble flow and would provide accurate radial distances for galaxies.
In reality, peculiar velocities are gravitationally induced by inhomogeneous structure and distort the measured distances. \citet{Kaiser:1987qv} described the anisotropy of the
clustering pattern in redshift space but restricted his calculation to large scales where linear 
perturbation theory should be applicable. In the linear regime, the matter power spectrum
in redshift space is a function of the power spectrum in real space and the parameter $\beta = f/b$ where $f$ is the linear growth rate. The 
linear bias factor, $b$, characterises the clustering of galaxies with respect to the underlying mass distribution \citep*[e.g.][]{Kaiser:1987qv}.
\citet{Scoccimarro:2004tg} extended the analysis of \citet{Kaiser:1987qv} into the non-linear regime, including the contribution of 
peculiar velocities on small scales. We test this model in this paper.

Perturbations in bulk flows converge more slowly then perturbations in density, and so very large volume simulations are needed to model these flows, and hence the redshift 
space distortion of clustering, accurately. Our simulation boxes are 125 times the volume of those used by \citet{Cole:1993kh}  and $\sim 30$ times the volume 
of the N-body results interpreted by \citet{Scoccimarro:2004tg}. \citet{2009MNRAS.393..297P} used a single $1h^{-1}$Gpc box to study redshift space
distortions in a $\Lambda$CDM model. Their simulation is over three time smaller than the one we consider.

This paper is organised as follows: In Section \ref{RSD} we discuss the linear growth rate and  review the
theory  of redshift space distortions  on linear and non-linear scales. In Section \ref{2} we present the quintessence models considered
 and the details of our N-body simulations. 
The main results of the paper are presented in Sections \ref{PS12} and \ref{RT12}. 
The linear theory redshift space distortion, as well as models  for the redshift space power spectrum which include non-linear effects are examined in Section \ref{PS12} 
for various dark energy cosmologies. In Section \ref{RT12} we present the density-velocity relation measured from the simulations. Using this relation the
 non-linear models used in the previous
section can be made
cosmology independent. We present a prescription for obtaining the non-linear velocity divergence power spectrum from the non-linear matter power spectrum at an 
arbitrary redshift in Section \ref{rt2}. Our conclusions are presented in Section \ref{5.1}.

\section{Redshift space distortions\label{RSD}}

In Section \ref{probeofgravity} we consider several parametrizations which are commonly used for the linear growth rate.
 In Section \ref{1.1} we review linear perturbation theory for redshift 
space distortions and discuss the assumptions that are used in this approach. In Section \ref{2.2} we present several models proposed to describe the distortions
in the non-linear regime. A similar review can be found in \citet{2009MNRAS.393..297P}.

\subsection{Linear growth rate as a probe of gravity \label{probeofgravity}}

The linear growth rate is a promising probe of the nature of dark energy
 \citep{ Guzzo:2008ac,Wang:2007ht,2008APh....29..336L, 2009JCAP...10..004S,2009MNRAS.397.1348W,2009MNRAS.393..297P,2010MNRAS.404..239S,2010PhRvD..81d3512S}.
Although the growth equation for dark matter perturbations is easy to solve exactly, it is common to consider parametrizations for 
the linear growth rate,
$f= {\rm{d ln } }D/{\rm{d ln}}a$, where $D(a)$ is the linear growth factor.
These parametrizations employ different variables with distinct dependencies on the expansion and growth histories.

A widely used approximation for $f$, first suggested by \citet{1976ApJ...205..318P}, is $f(z) \approx \Omega_{\rm m}^{0.6}$.
\citet{1991MNRAS.251..128L} found an expression for $f$, 
in terms of the present day densities of matter, $\Omega_{\rm m}$, and dark energy,
 $\Omega_{\tiny \mbox{DE}}$,
 which showed only a weak dependence on the dark energy density,
with $f \approx \Omega_{\rm m}^{0.6} + \Omega_{\tiny \mbox{DE}}/70 \left( 1 + \Omega_{\rm m}/2 \right)$.
\citet{Linder:2005in} extended the analysis of \citet{Wang:1998gt} to find a new fitting formula to the exact solution for
the growth factor, which he cast in the following form
\begin{equation}
\label{linder}
g(a) = \frac{D(a)}{a} \approx \mbox{exp} \left(\int^a_0 {\rm d} { \ln}a \, [\Omega^{\gamma}_{\rm m}(a) -1]\right) \, ,
\end{equation}
where $\gamma$ is the index which parametrises the growth history, while the expansion
history is described by the matter density $\Omega_{\rm m}(a)$.
\citet{Linder:2005in} proposed the empirical result
 $\gamma = 0.55 +0.05[1+w(z=1)]$, where $w$ is the dark energy equation of state, which gives $f = \Omega_{\rm m}^{0.55}$ for a cosmological constant
\citep*[see also][]{2007APh....28..481L}.
We discuss this formula for $f$ further in Section \ref{2} when we introduce the quintessence dark energy models used in this paper.

\subsection{Linear redshift space distortions \label{1.1}}

The comoving distance to a galaxy, $\vec{s}$,  differs from its true distance, $\vec{x}$, due to its peculiar velocity, $\vec{v}(\vec{x})$
(i.e. an additional velocity to the Hubble flow), as
\begin{equation}
s = x + \frac{\vec{v} \cdot \hat{x}}{H(a)},
\end{equation}
where $H(a)$ is the Hubble parameter and $\vec{v} \cdot \hat{x}$ is the peculiar velocity along the line of sight. 
Inhomogeneous structure in the universe induces peculiar motions which distorts the clustering
pattern measured in redshift space on all scales. This effect must be taken into account when analyzing three dimensional datasets which use
redshift as the radial coordinate.
Redshift space effects alter the appearance of the clustering
of matter, and together with non-linear evolution and bias, lead the power spectrum to depart from simple linear perturbation theory predictions.

On small scales, randomised velocities associated with  viralised structures decrease the power.
The dense central regions of galaxy clusters look elongated along the line of sight in redshift space, which produces \lq fingers
of God\rq \, \citep{Jackson} in redshift survey cone plots.
On large scales, coherent bulk flows distort  clustering
 statistics, \citep[see][for a review of redshift space distortions]{1998ASSL..231..185H}.
For growing perturbations on large scales, the overall effect of redshift space distortions is to enhance the clustering amplitude.
Any difference in the velocity field due to mass flowing from underdense regions to high density regions will alter the volume element, causing
an enhancement of the apparent density contrast in redshift space, $\delta_s(\vec{r})$, compared to that in real space, $\delta_r(\vec{r})$.
This  effect was first analyzed by \citet{Kaiser:1987qv} and can be approximated by
\begin{equation}
\label{k}
\delta_s(r) = \delta_r(r)(1+\mu^2 \beta) ,
\end{equation}
where $\mu$ is the cosine of the angle between the wavevector, $\vec{k}$, and the line of sight, $\beta =  f/b$ 
 and  the bias, $b=1$ for dark matter.

The Kaiser formula (Eq. \ref{k}) relates the overdensity in redshift space to the corresponding value in real space using several approximations:

1. The small scale velocity dispersion can be neglected.

2. The velocity gradient $|\rm{d}\vec{u}/\rm{d}r| \ll 1$.

3. The velocity and density perturbations satisfy the linear continuity equation.

4. The real space density perturbation is assumed to be small, $|\delta(r)| \ll 1$, so that higher
order terms can be neglected.

All of these assumptions are valid on scales that are well within the linear regime
and will break down on different scales as the density fluctuations grow. The linear regime is therefore defined over a different range of scales for each effect.

The matter power spectrum in redshift space can be decomposed into multipole moments using Legendre polynomials, $L_l(\mu)$,
\begin{eqnarray}
P(k,\mu) = \sum_{l=0}^{2l} P_l(k) L_l(\mu) \, .
\end{eqnarray}
The anisotropy in $P( \vec{k} )$ is symmetric in $\mu$, as $P(k,\mu)=P(k,-\mu)$, so only even values of $l$ are summed over. Each multipole moment is given by
\begin{eqnarray}
P^s_l(k) = \frac{2l+1}{2} \int_{-1}^{1} P(k,\mu) L_l(\mu) \rm{d}\mu \, ,
\end{eqnarray}
where the first two non-zero moments have Legendre polynomials, $L_0(\mu) = 1$ and  $L_2(\mu) = (3\mu^2 - 1)/2$.
Using the redshift space density contrast, Eq. \ref{k} can be used to form $P(k,\mu)$ and then integrating over the 
cosine of the angle $\mu$ gives the 
 spherically averaged monopole power spectrum in redshift space, $P^s_0(k)$,
\begin{eqnarray}
\label{mr}
\frac{P^s_0(k)}{P^r(k)} = 1 + \frac{2}{3}f + \frac{1}{5}f^2 \, ,
\end{eqnarray}
where $P^r(k)$ denotes the matter power spectrum in real space.
In practice, $P^r(k)$ cannot be obtained directly for a real survey without making approximations \citep[e.g.][]{1994MNRAS.270..183B}.

\begin{figure*}
{\epsfxsize=13.5truecm
\epsfbox[150 463 480 683 ]{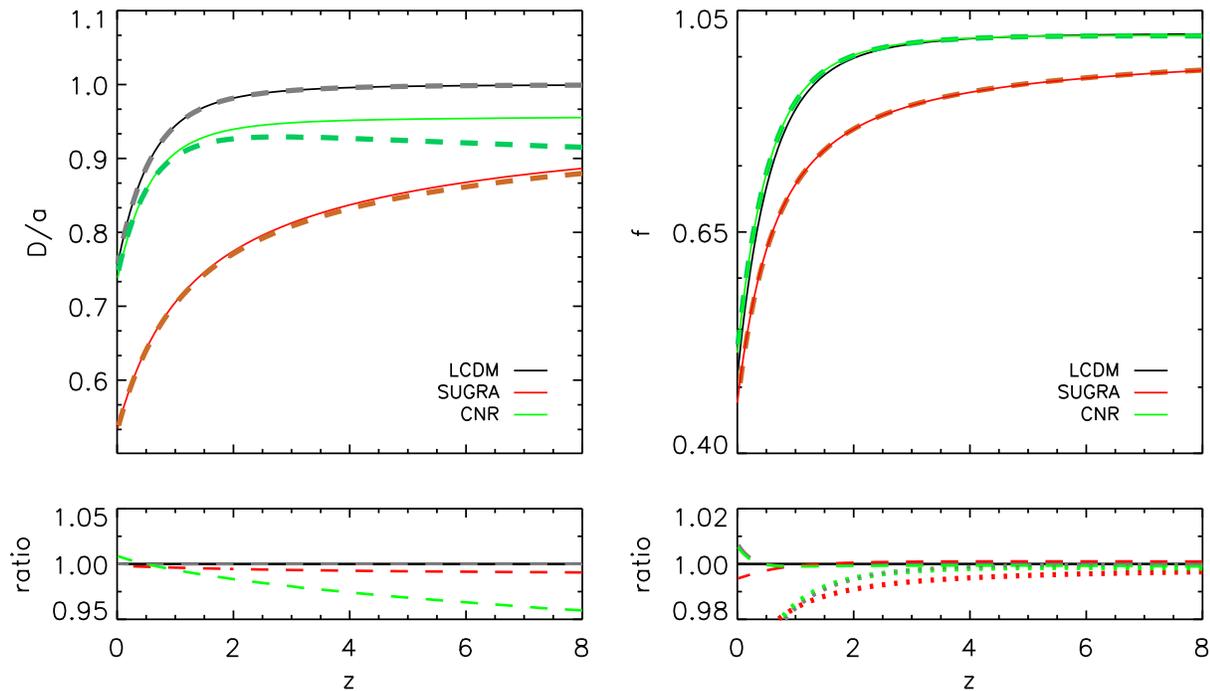}}
\caption{Left panel:
The linear growth factor divided by the scale factor as a function of redshift for the SUGRA and CNR quintessence models and $\Lambda$CDM, as indicated by the key.
Right panel:
The linear growth rate, $f = {\rm{d}} {\rm{ln}} D/{\rm{d ln}} a$, for the two dark energy models and $\Lambda$CDM as a function of redshift.
In both the left and right main panels, solid lines represent the exact solution
for the linear growth factor and growth rate and dashed lines show the fitting formula given in Eq. \ref{linder}.
 Note in the right main panel the $\Lambda$CDM grey dashed line has been omitted for clarity.
The lower left hand panel shows the formula for $D(a)/a$ given by \citet{Linder:2005in} divided by the exact solution as a function of redshift.
The ratio of the formula in Eq. \ref{linder}  for the growth
rate, $f$, to the exact solution is shown in the lower right hand panel.
Also in the lower right panel the dotted lines show the ratio of the fitting formula $f= \Omega^{0.6}_{\rm m}$ to the exact solution for each of the dark energy
models plotted as a function of redshift.
} \label{gf}
\end{figure*}

In this paper we also consider the estimator for $f$ suggested by \citet{Cole:1993kh}, which is the ratio of quadrupole to monopole moments  of the redshift 
space power spectrum, $P_2^s(k)/P_0^s(k)$.
From Eq. \ref{k} and after spherically averaging, the estimator for $f$ is then
\begin{eqnarray}
\label{dmm}
\frac{P_2^s(k)}{P_0^s(k)} = \frac{4f/3 +4f^2/7}{1 + 2f/3 + f^2/5} \, ,
\end{eqnarray}
which is independent of the real space power spectrum. Here, as before, $f = \beta/b$, with $b=1$ for dark matter.


\subsection{Modelling  non-linear distortions to the power spectrum in redshift space \label{2.2} }

Assuming the line of sight component is along the $z$-axis, the  fully non-linear relation between the real and redshift space power spectrum can be written as
\citep{Scoccimarro:1999ed}
\begin{eqnarray}
\label{NL}
P^s(k,\mu) =  \int \frac{\rm{d}^3 \bf{r}}{(2\pi)^3} e^{-i \bf{k} \cdot \bf{r}} \langle e^{i\lambda \Delta u_z} [\delta({\bf{x}}) - f \nabla_z \cdot u_z({\bf{x}})] \nonumber  \\
 \times [\delta({\bf{x'}}) - f \nabla_z' \cdot u_z({\bf{x'}})]\rangle \, ,
\end{eqnarray}
where $\lambda = f k \mu$, $u_z$ is the comoving peculiar velocity
along the line of sight, $\Delta u_z = u_z({\bf{x}}) - u_z(\bf{x'})$,  $\bf{r} = \bf{x} -\bf{x'}$
 and the only approximation made is the plane parallel approximation. This expression is the Fourier analog of the \lq streaming model\rq \, first suggested by
\citet{1980:Peebles} and modified by \citet{1995ApJ...448..494F}           to take into account the density-velocity
coupling.
At small scales (as $k$ increases) the exponential component damps the power, representing the impact of randomised velocities inside gravitationally 
bound structures.

Simplified models for redshift space distortions are frequently used. 
Examples include multiplying Eq. \ref{mr} by a factor which
attempts to take into account small scale effects and  is either a Gaussian or an exponential \citep{Peacock:1993xg}.
A popular phenomenological example of this which incorporates the damping effect of velocity dispersion on small scales is the so called \lq dispersion model\rq \,
\citep{Peacock:1993xg},
\begin{eqnarray}
P^s(k,\mu) =  P^r(k) (1+\beta \mu^2)^2 \frac{1}{(1 + k^2 \mu^2 \sigma_p^2/2)} \, ,
\end{eqnarray}
where $\sigma_p$ is the pairwise velocity dispersion along the line of sight, which is treated as a parameter to be fitted to the data. 
Using numerical simulations, \citet{1999MNRAS.310.1137H}
found a fit to the quadrupole to monopole ratio 
$P^s_2/P^s_0=(P^s_2/P^s_0)_{\tiny \mbox{lin}}(1-x^{1.22}) $ to mimic damping and non-linear effects, 
where $(P^s_2/P^s_0)_{\tiny \mbox{lin}}$ is the linear theory prediction given by Eq. \ref{dmm},  $x=k/k_1$ and $k_1$ is a free parameter.
They extended the dynamic range of simulations, to replicate the effect of a larger box, using the 
approximate method for adding long wavelength power suggested by \citet{1997MNRAS.286...38C}. 

The velocity divergence auto power spectrum is the emsemble average, $P_{\theta \theta} = \langle |\theta|^2 \rangle $ where $\theta = \vec{\nabla}\cdot \vec{u}$ is the velocity
divergence. The cross power spectrum of the velocity divergence and matter density is $P_{\delta \theta}= \langle |\delta \theta| \rangle$,
where in this notation the matter density auto spectrum is $P_{\delta \delta}= \langle |\delta|^2 \rangle$.
In Eq. \ref{NL}, the term in square brackets can be re-written in terms of these non-linear velocity divergence power spectra  by multiplying out the
 brackets and using the fact that $\mu_i = \vec{k}_i \cdot
\hat{z}/k_i$.
\citet{Scoccimarro:2004tg} proposed the following model for the redshift space power spectrum in terms of $P_{\delta \delta}$, the non-linear
matter power spectrum,
 $P_{\theta \theta}$ and $P_{\delta \theta}$,
\begin{eqnarray}
\label{SM}
&&P^s(k,\mu)=\\  \nonumber
&&  \left( P_{\delta \delta}(k) + 2 f\mu^2 P_{\delta \theta}(k) + f^2\mu^4P_{\theta \theta}(k)\right) 
\times e^{-(f k \mu \sigma_v )^2} \, ,
\end{eqnarray}
where   $\sigma_v$ is the 1D linear velocity dispersion given by
\begin{eqnarray}
 \sigma^2_v = \frac{1}{3}\int\frac{P_{\theta \theta}(k)}{k^2} {\rm d}^3k. 
\end{eqnarray}
In linear theory, $P_{\theta \theta}$ and $P_{\delta \theta}$ take the same form as $P_{\delta \delta}$ and depart from this at different scales.
Using  a simulation  with 512$^3$ particles in a box of length $479 h^{-1}$Mpc \citep{Yoshida:2001vf},
\citet{Scoccimarro:2004tg} showed that this simple ansatz  for $P_s(k,\mu)$ was an improvement over the Kaiser formula when comparing to N-body simulations in a $\Lambda$CDM
cosmology. 
As this is a much smaller simulation volume than the one we  use to investigate  redshift space distortions we are able to test the fit to the 
measured power spectrum
on much larger scales and to higher accuracy.

\section{N-body simulations of dark energy \label{2}}

In the following sections we briefly review the quintessence models discussed in this paper and the N-body simulations used to measure various power spectra.

\begin{figure*}
{\epsfxsize=18.truecm
\epsfbox[84 448 496 596]{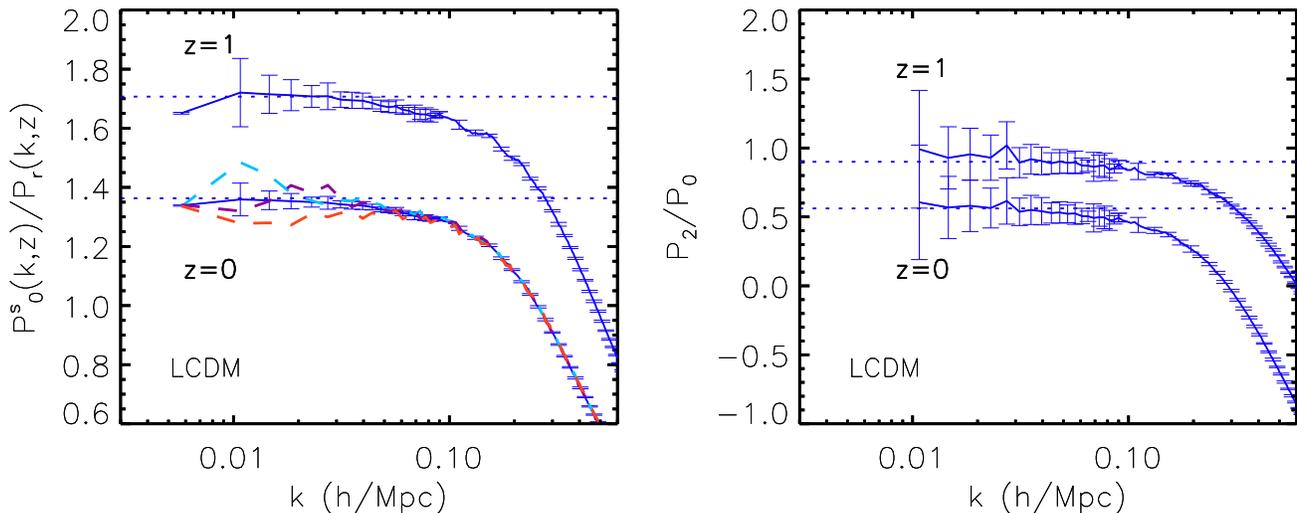}}
 \caption{Left panel: The ratio of the monopole redshift power spectra  and real space power spectra measured from the $\Lambda$CDM simulation
 at $z=0$ and $z=1$ are plotted
as blue lines.
The error bars plotted represent the scatter between the different power spectra from four $\Lambda$CDM
simulations
 set up with different realisations of the density field with the distortions imposed along either the $x, y$ or $z$ axis and averaged.
The power spectra $P(k,\mu=k_x/k)$, $P(k,\mu=k_y/k)$ and $P(k,\mu=k_z/k)$ measured from one simulation are plotted as the cyan, purple and red dashed lines respectively.
Right panel: The ratio of the quadrupole to monopole moment of the redshift space power spectrum measured from the simulations at $z=0$ and $z=1$ in $\Lambda$CDM are plotted
in blue.
It was not possible to accurately measure the quadrupole to monopole power in the first bin, so this point has not been plotted in the right hand panel.
Note for wavenumbers $k>0.1h$Mpc$^{-1}$, only every fifth error bar is plotted for clarity.
The Kaiser formula, given by Eq. \ref{mr}, is plotted as a blue dotted line. The error bars were obtained as described for the left-hand panel.
}\label{lcdm}
\end{figure*}

\subsection[]{Quintessence models  \label{QUIN}}

In quintessence models of dark energy, the cosmological constant  is replaced by an extremely light scalar field
which evolves slowly \citep{Ratra:1987rm,1988NuPhB.302..668W,1998PhRvL..80.1582C,Ferreira:1997hj}.
Different quintessence dark energy models have different dark energy densities as a function of time, $\Omega_{\tiny \mbox{DE}}(z)$.
This implies a different growth history for dark matter perturbations from that expected in $\Lambda$CDM.
In this paper we consider three quintessence models, each with a different evolution for the dark energy equation of state parameter, $w(a)$. These models 
are a representative sample of a range of quintessence models and are a subset  of those considered by
\citet{2010MNRAS.401.2181J} to which we refer the reader for further details. Briefly, the SUGRA 
model of \citet{Brax:1999gp} has an equation of state today of  $w_0 =-0.82$ and a linear growth factor which differs from $\Lambda$CDM by $~20\%$ at $z=5$.
The 2EXP model has an equation of state that makes a rapid transition to $w_0 =-1$ at $z=4$  and since then 
has a similar expansion history to $\Lambda$CDM \citep{2000PhRvD..61l7301B}.
The CNR quintessence model  has a non-negligible amount of dark energy at early times and an equation of state today of $w_0=-1$  \citep{Copeland:2000vh}.
The dark energy equation of state for each model is described using a
4 variable parametrization for $w(a)$ which is able to accurately describe the expansion history over the full range of redshifts modelled by the simulations 
\citep{Corasaniti:2002vg}.

The presence of small but appreciable amounts of dark energy at early times also modifies the growth
rate of fluctuations from that expected in a matter dominated universe and hence changes the shape of the linear theory $P(k)$ from the $\Lambda$CDM prediction 
\citep{2010MNRAS.401.2181J}.
The CNR quintessence model used in this paper has non-negligible amounts of dark energy at high redshifts and so could be classed as an  \lq early
 dark energy\rq \, model \citep{Doran:2006kp}. As a result, the linear theory power spectrum is appreciably 
different from that in a $\Lambda$CDM cosmology, with a broader turnover,
 \citep[see][for futher details]{2010MNRAS.401.2181J}.

Quintessence dark energy models will not necessarily
agree with observational data if we adopt
the same cosmological parameters as used in the best fitting $\Lambda$CDM
cosmology.
These best fit parameters were found using the  observational
 constraints on distances such as the angular diameter distance to last scattering and the sound horizon at this epoch,
from the cosmic microwave background, as well as distance measurements from the baryonic 
acoustic oscillations and Type Ia supernovae \citep{2010MNRAS.401.2181J}.
In this paper
 the best fitting cosmological parameters for each quintessence model are used in the N-body
simulations, as listed in Table 1.

\begin{table}
\label{table}
\caption{Cosmological parameters used in the simulations. The first column gives the cosmological model, the second the present day  matter density, $\Omega_{\rm m}$,
the third the baryon density, $\Omega_{\rm b}$ and the fourth the  Hubble constant, $h$, in units of 100 km s$^{-1}$ Mpc$^{-1}$.
}
\begin{tabular}{|c|r|r|r|}
\hline
Model & $\Omega_{\rm m}$ & $\Omega_{\rm b}$ & $h$ \\
\hline
$\Lambda$CDM / 2EXP& 0.26 & 0.044 &  0.715  \\
SUGRA  & 0.24  & 0.058  &  0.676\\
CNR & 0.28& 0.042 & 0.701\\
\hline
\end{tabular}
\end{table}

In the left panel of Fig. \ref{gf}, we plot the exact solution for the linear theory growth factor, divided by the scale factor,
as a function of redshift together with the fitting formula in Eq. \ref{linder}.
The 2EXP quintessence model is not plotted in Fig. \ref{gf} as the linear growth factor for this model differs from $\Lambda$CDM only at high redshifts, $z>10$.
\citet{Linder:2005in} found that the formula in Eq. \ref{linder} reproduces the growth factor to better than 0.05\%
for $\Lambda$CDM cosmologies and to $\sim 0.25$\% for different dynamical quintessence models to the ones considered in this paper.
We have verified that this fitting formula for $D$ is accurate to $\sim 1\%$ for the SUGRA and 2EXP dark energy models used in this paper,
over a range of redshifts.
Note, in cosmological models which feature non negligible amounts of dark energy at high redshifts,
a further correction factor is needed to this parametrisation \citep{Linder:2009kq}. 
Using the parametrization for $w(a)$ provided by \citet{Doran:2006kp} for \lq early dark energy\rq,  \citet{Linder:2009kq} proposed
a single correction factor  which was independent of redshift.
The CNR model has a high fractional dark energy density at early times and as a result we do not expect the linear theory growth to be accurately reproduced by Eq. \ref{linder}.
As can be seen in Fig. \ref{gf} for the CNR model,
 any correction factor between the fitting formula suggested by \citet{Linder:2005in} and the exact solution for $D/a$
would depend on redshift and is not simply a constant. In this case, the \lq early dark energy\rq \,parametrisation 
of \citet{Doran:2006kp} is not accurate enough to fully describe 
the dynamics of the CNR quintessence model.
This difference is $\sim$5\% at $z=8$ for the CNR model, as can be seen in the ratio plot in the left panel of Fig \ref{gf}.
The exact solution for the linear growth rate, $f$, and the fitting formula in Eq. \ref{linder}, $f = \Omega^{\gamma}_{\rm{m}}(a)$,
is plotted in the right panel of Fig. \ref{gf}.
The old approximation  $f= \Omega^{0.6}_{\rm m}$, is plotted  in the bottom right panel in Fig. \ref{gf}. 
The dotted lines represent the ratio  $f = \Omega^{0.6}_{\rm m}$ to the exact solution for each of the dark energy
models. It is clear that this approximation for the growth factor is not as accurate as the formula in Eq. \ref{linder} over the same range of redshifts.

\subsection{Simulation Details}

 We use the N-body simulations carried out by \citet{2010MNRAS.401.2181J}.
These simulations were performed at the Institute of Computational Cosmology using a memory efficient version of the  TreePM
  code  {\sc  Gadget-2}, called {\sc L-Gadget-2} \citep{Springel:2005mi}. For the $\Lambda$CDM model we used
the following cosmological parameters:  
$\Omega_{\rm m} = 0.26$,
 $\Omega_{\rmn{DE}}=0.74$, $\Omega_{\rm b} = 0.044$,
$h = 0.715$ and a spectral tilt of $n_{\mbox{s}} =0.96$ \citep{2009MNRAS.400.1643S}. 
The  linear theory rms fluctuation 
in spheres of radius 8 $h^{-1}$ Mpc is set to be  $\sigma_8 = 0.8$. 
For each of the quintessence models, a four variable parametrization 
of the dark energy equation of state is used as described above.
In each case, the cosmological parameters used are the best fitting parameters
to observational constraints from the cosmic microwave background, baryonic acoustic oscillations and supernovae Ia 
taking into account the impact of the quintessence model. \citep[Stage III in the terminology of ][]{2010MNRAS.401.2181J}.

The simulations use $N=646^3 \sim 269 \times 10^6$ particles to represent the  matter distribution in a  computational box of 
comoving length $1500 h^{-1}$Mpc. 
 The comoving softening length is $50 h^{-1}$kpc. 
The particle mass in the $\Lambda$CDM simulation is $9.02 \times 10^{11}  h^{-1}
M_{\sun}$
and is slightly different in the other runs due to changes in $\Omega_{\rm{m}}$ (see Table 1).
The initial conditions were set up starting from a 
 glass configuration of particles 
 \citep{1994RvMA....7..255W,Baugh:1995hv}. 
In order to limit the impact of the initial displacement scheme
 we chose a starting redshift of $z=200$. 

The linear theory power spectrum used to generate the initial 
conditions was  obtained using CAMB \citep{Lewis:2002ah}.  
 We use a modified version of CAMB which incorporates 
the influence of dark energy on dark matter clustering at early times \citep{Fang:2008sn}. 

 In each model the power spectrum at redshift zero is normalised to have 
$\sigma_8 = 0.8$. Using the linear growth factor for each dark energy model, the 
linear theory $P(k)$ was then evolved backwards to the 
starting redshift of $z=200$ in order to generate the initial 
conditions. 
The power spectrum was computed by assigning the particles to a mesh using the cloud in cell (CIC) assignment scheme \citep{1981csup.book.....H} and 
performing a fast Fourier transform (FFT) of the density field.
To compensate for the mass assignment scheme we perform an approximate de-convolution following  \citet{1991ApJ...375...25B}.

\begin{figure*}
{\epsfxsize=18.truecm
\epsfbox[84 448 496 596]{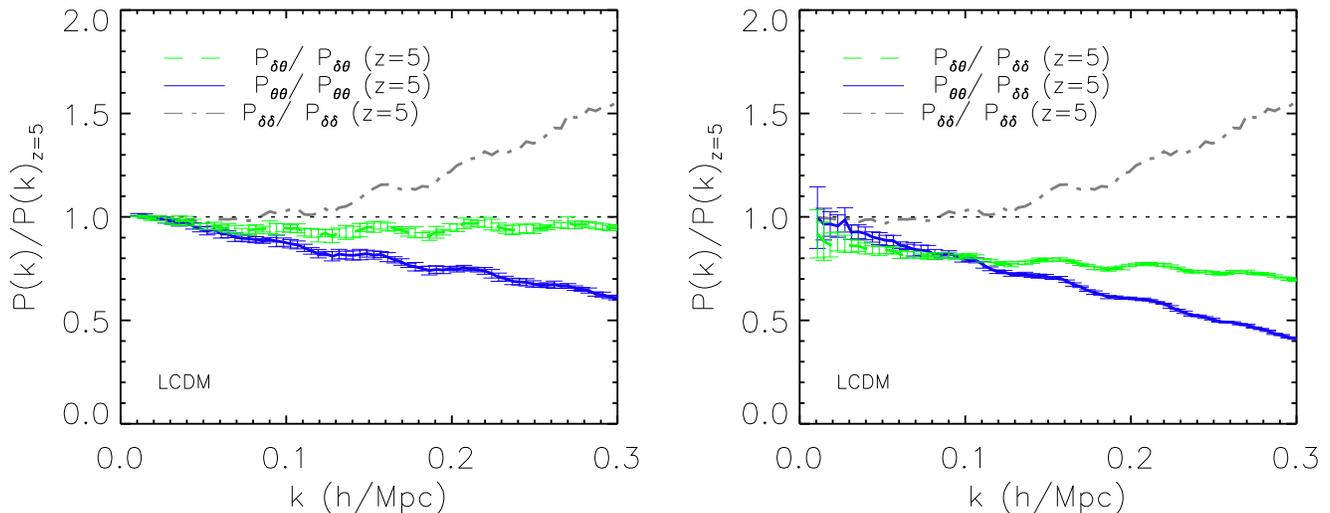}}
\caption{
Left panel:
 The ratio of the non-linear power spectra, $P_{\delta \delta}$,  $P_{\delta \theta}$ and $P_{\theta \theta}$ for  $\Lambda$CDM measured
from the simulation at $z=0$,
divided by the corresponding power spectrum
measured from the simulation at $z=5$, scaled using the square of the ratio of the linear growth factor at $z=5$ and $z=0$.
The non-linear matter power spectrum is  plotted as a grey dot-dashed line, the non-linear velocity divergence auto power spectrum  $P_{\theta \theta}$ is plotted as a
blue solid line and the non-linear cross power spectrum, $P_{\delta \theta}$, is plotted as a green dashed line.
Right panel:  The ratio of the non-linear power spectra, $P_{\delta \delta}$, $P_{\delta \theta}$ and $P_{\theta \theta}$,
 to the linear theory matter $P(k)$  in $\Lambda$CDM measured from the simulation at $z=0$.
All power spectra have been divided by the
linear theory matter power spectrum measured from the simulation at $z=5$, scaled using the square of the ratio of the linear growth factor at $z=5$ and $z=0$.
In both panels the error bars represent the scatter  over  eight $\Lambda$CDM realisations after imposing the
peculiar velocity distortion along each cartesian axis in turn.
}\label{ratio}
\end{figure*}

\begin{figure*}
{\epsfxsize=18.truecm
\epsfbox[84 448 496 596]{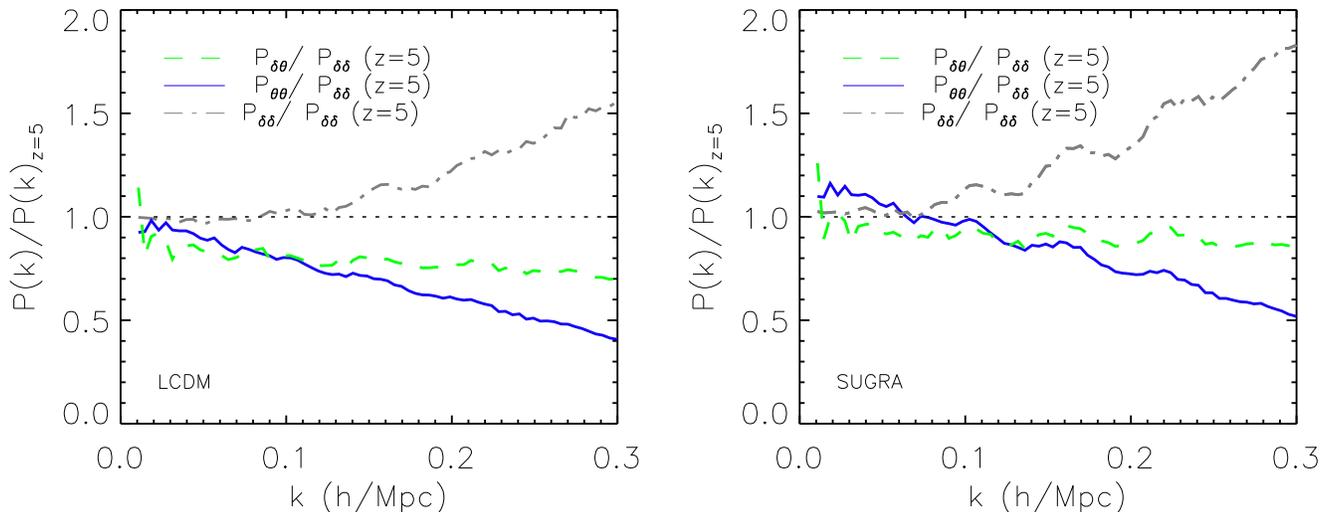}}
\caption{
 Left panel: The ratio of the non-linear power spectra, $P_{\delta \delta}$, $P_{\delta \theta}$ and $P_{\theta \theta}$,
 to the linear theory $P(k)$  in $\Lambda$CDM measured from one realisation of the matter density and velocity fields at $z=0$.
All power spectra have been divided by the
linear theory matter power spectrum measured from the simulation at $z=5$, scaled using the square of the ratio of the linear growth factor at $z=5$ and $z=0$.
Right panel: Similar  to that in the left panel but for the SUGRA quintessence model.
The lines are the same as used in the left hand panel.
}\label{ratio2}
\end{figure*}

\section{Results I: the matter power spectrum in real and redshift space \label{PS12} }

In Sections \ref{4.1} and \ref{4.2} we present  the redshift space distortions measured from the simulations in $\Lambda$CDM and quintessence cosmologies, and we 
compare with the predictions of the linear and non-linear models discussed in Section \ref{2.2}. 

\subsection{Testing the linear theory redshift space distortion\label{4.1}}

In the left panel of Fig. \ref{lcdm}, we plot the ratio of the redshift space to real space power spectra, 
measured from the  $\Lambda$CDM simulation at $z=0$ and $z=1$.
Using the  plane parallel approximation, we assume the observer is at infinity and as a result the velocity distortions are imposed along one direction in $k$-space. 
If we choose the line of sight direction to be the 
$z$-axis, for example, then $\mu = k_z/k$ where $k =|\vec{k}|$. In this paper the power spectrum in redshift space represents the average of $P(k,\mu=k_x/k)$, 
 $P(k,\mu=k_y/k)$ and  $P(k,\mu=k_z/k)$ where the line of sight components are parallel to the $x$, $y$ and $z$ directions respectively. 
 We use this average as there is a significant scatter in the amplitudes of the three redshift space power spectra on large scales, even for a computational box 
as large as the one we have used.
The three monopoles of the redshift space power spectra $P(k,\mu=k_x/k)$, $P(k,\mu=k_y/k)$ and $P(k,\mu=k_z/k)$ measured in one of the realisations are plotted as
the cyan, purple and red dashed lines respectively, to illustrate the scatter.


In Fig. \ref{lcdm}  the Kaiser formula, given by Eq. \ref{mr}, is plotted as a blue dotted line, using a value of $f = \Omega^{0.55}_m(z)$ for $\Lambda$CDM.
The error bars plotted represent the scatter 
over four realisations after averaging over  
$P(k)$ obtained by treating the 
$x, y$ and $z$ directions as the line of sight.
 It is clear from this plot that the linear perturbation theory limit is only attained on extremely large scales $(k <0.03 h$Mpc$^{-1}$$)$ at $z=0$ and at $z=1$.
Non-linear
effects are significant on scales $0.03<k (h$Mpc$^{-1}$$)<0.1$ which are usually considered to be in the linear regime.
The measured variance in the matter power spectrum on these scales is $10^{-3}<\sigma^2 <10^{-2}$.

 In the right panel of Fig. 2 we plot the ratio $P^s_2/P^s_0$ for
$\Lambda$CDM at $z=0$ and $z=1$. The ratio agrees with the Kaiser
limit (given in Eq. 7) down to smaller scales, $k<0.06 h$Mpc$^{-1}$,
compared to the monopole ratio plotted in the left panel.
Our results agree with previous work 
on the quadrupole and monopole moments of the redshift space power spectrum for $\Lambda$CDM \citep{Cole:1993kh,1999MNRAS.310.1137H,Scoccimarro:2004tg}.
At $z=1$, the damping effects are less prominent and the Kaiser limit is attained over a slightly wider range of  scales, $k < 0.1 h$Mpc$^{-1}$, as non-linear 
effects are smaller then at $z=0$.
In the next section, we consider these ratios for the quintessence dark energy models in more detail. 
For each model we find that the analytic expression for the quadrupole to monopole ratio 
describes the simulation results over a wider range of wavenumber then
 the analogous result for the monopole moment.

\begin{figure}
{\epsfxsize=7.truecm
\epsfbox[96 388 260 702]{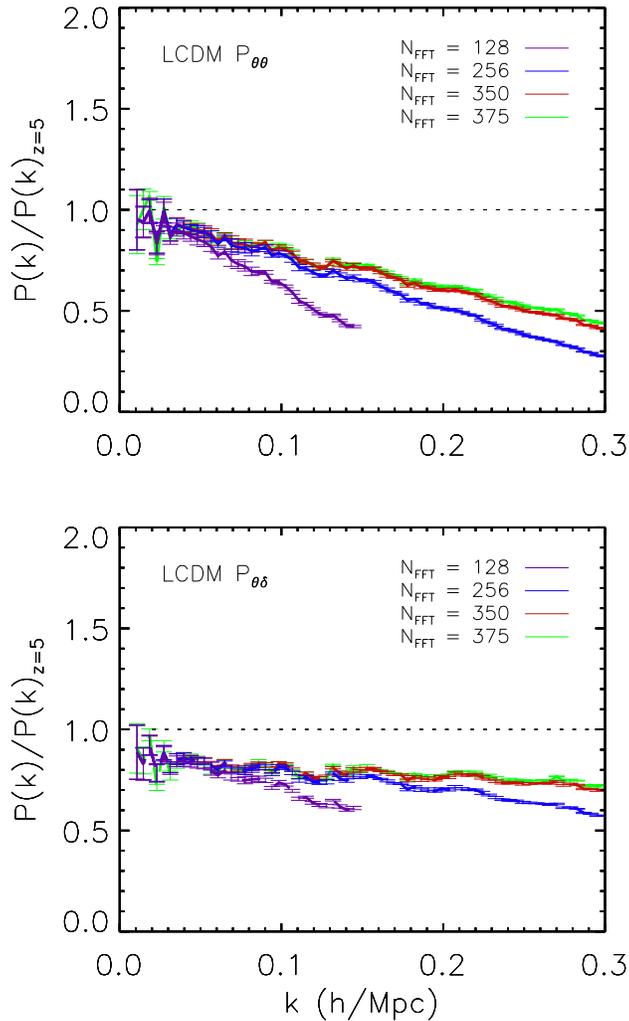}}
\caption{
A comparison of the impact of the FFT grid dimension on power spectrum estimation. The plots show
the ratio of the non-linear power spectra,  $P_{\theta \theta}$ (upper panel) and $P_{\delta \theta}$ (lower panel),
 to the linear theory matter power spectrum measured from the simulations  in $\Lambda$CDM,
 using different FFT grid sizes. From bottom to top in each panel the lines show the ratios for grid sizes
$N_{\tiny\mbox{FFT}} =128$ (purple), $N_{\tiny\mbox{FFT}} =256$ (blue), $N_{\tiny\mbox{FFT}} =350$ (red) and $N_{\tiny\mbox{FFT}} =375$ (green).
}\label{ratio128}
\end{figure}

\subsection{Nonlinear models of $P_s(k,\mu)$ \label{4.2} }

The linear theory relationship between the real and redshift space power spectra given in Eq. \ref{mr} assumes various non-linear effects are small and can be neglected on large scales. These assumptions are listed in Section \ref{1.1}. 
In this section we consider the  
non-linear terms in the gradient of the line of sight velocity field and explore the scales at which it is correct to  ignore such effects in the redshift space power spectrum.
As a first step, we compare the model in Eq. \ref{SM}, to measurements from N-body simulations for different quintessence dark energy models, without the damping term due to 
velocity dispersion. This will highlight the scale at which non-linear velocity divergence terms affect the matter power spectrum in redshift space and cause it to depart from 
the linear theory prediction.

If we rewrite $\rm{d}\delta/\rm{d}\tau$ as $a H(a) f(\Omega_{\rm m}(a), \gamma)\, \delta$, where $\delta$ is the matter perturbation and $\tau$ is 
the conformal time, $\rm{d}t = a(\tau)\rm{d}\tau $, then the linear continuity equation becomes
\begin{eqnarray}
\label{con}
\theta = \vec{\nabla}\cdot \vec{u} = -aH f \delta \, .
\end{eqnarray}
Throughout this paper we normalise the velocity divergence as $\theta(k,a) /[-a H(a) f(\Omega_{\rm m}(a),\gamma)]$, so $ \theta = \delta$ in the linear regime.
The volume weighted velocity divergence power spectrum is calculated from the simulations according to the prescription given in \citet{Scoccimarro:2004tg}. 
We interpolate the velocites and the densities onto a grid of $350^3$ points and then 
measure the ratio of the interpolated momentum to the interpolated density field. In this way, we avoid having to correct for the CIC assignment scheme. 
A larger grid dimension could result in empty cells where $\delta \rightarrow 0$.
A FFT grid of $350^3$ was used to ensure all grid points 
had non-zero density and hence a well defined velocity 
at each point.  
We only plot  the velocity power spectra in each of the figures up to half the 
Nyquist frequency for our default choice of $N_{\mbox{\tiny {FFT}}} = 350^3$, $k_{nq}/2 = \pi N_{\mbox{\tiny {FFT} }}/(2 L_{\tiny\mbox{box}} ) = 0.37 h$Mpc$^{-1}$ which is  beyond the range typically used in BAO fitting
when assuming linear theory.
\begin{figure*}
{\epsfxsize=15.5truecm
\epsfbox[131 380 469 705]{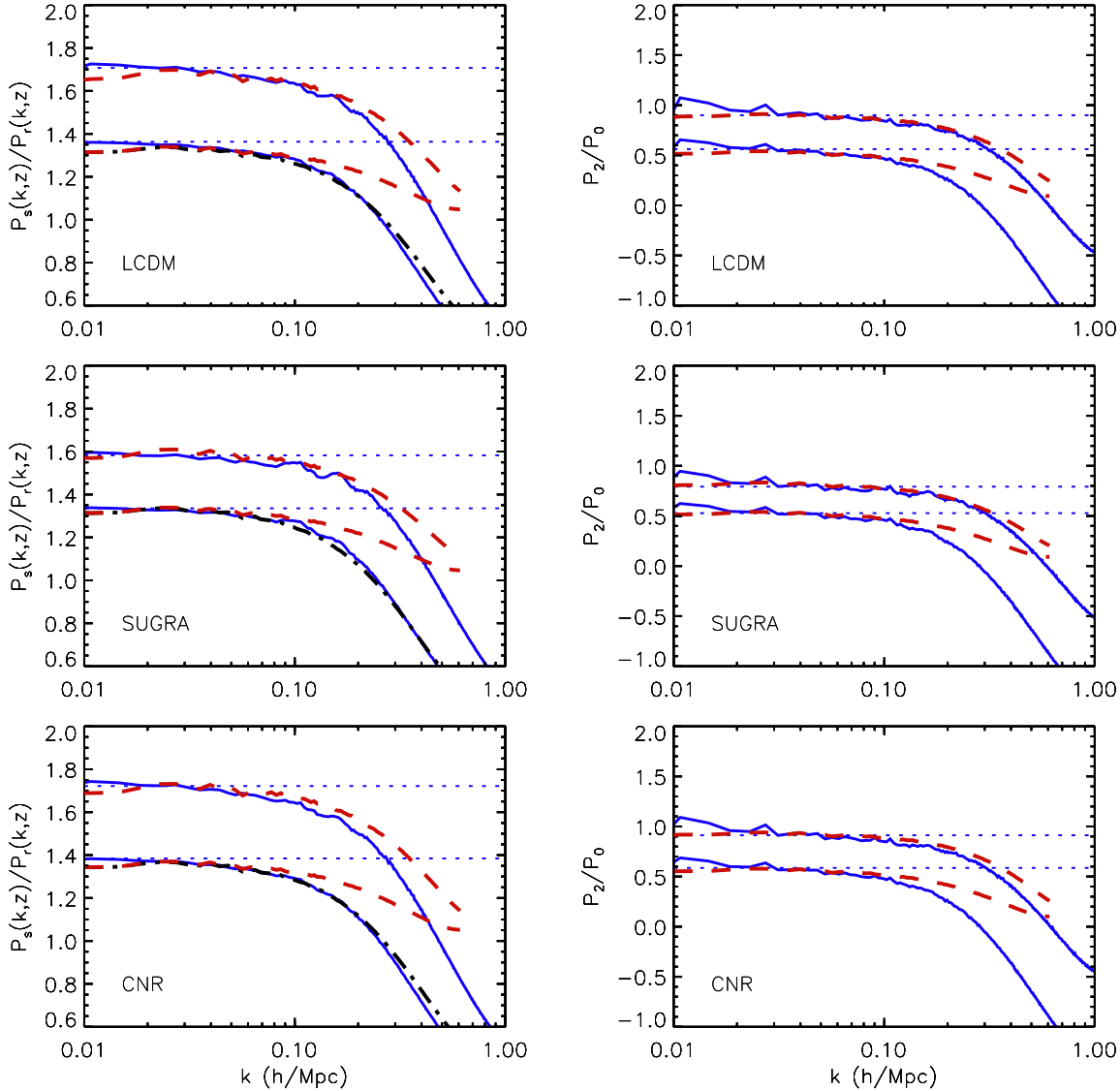}}
 \caption{
The left hand column shows the ratio of the  monopole of redshift power spectra to the real space power spectra at $z=0$ and $z=1$.
The right hand column shows the ratio of the quadrupole to monopole moment of the redshift space power spectra at $z=0$ and $z=1$.
 Different rows show different dark energy models as labelled.
Top row: The ratio of the redshift and real space power spectra  in $\Lambda$CDM are plotted
as solid lines in the left panel.
The dashed lines represent the same ratio using Eq. \ref{10} for the monopole of the redshift space power spectrum. The  dot-dash line represents the
 model given in Eq.
\ref{SM} which includes velocity dispersion effects.
In the right panel the ratio of the quadrupole to monopole moment of the redshift space power spectra  in $\Lambda$CDM are plotted
as  solid lines.
 The same ratio using Eq. \ref{dm} for the redshift space power spectrum is plotted as  dashed lines.
Middle row: Same as the top row but for the SUGRA quintessence model.
Bottom row: Same as the middle row but for the CNR quintessence model.
}\label{sugracnr}
\end{figure*}

\begin{figure*}
{\epsfxsize=16.5truecm
\epsfbox[79 442 546 685]{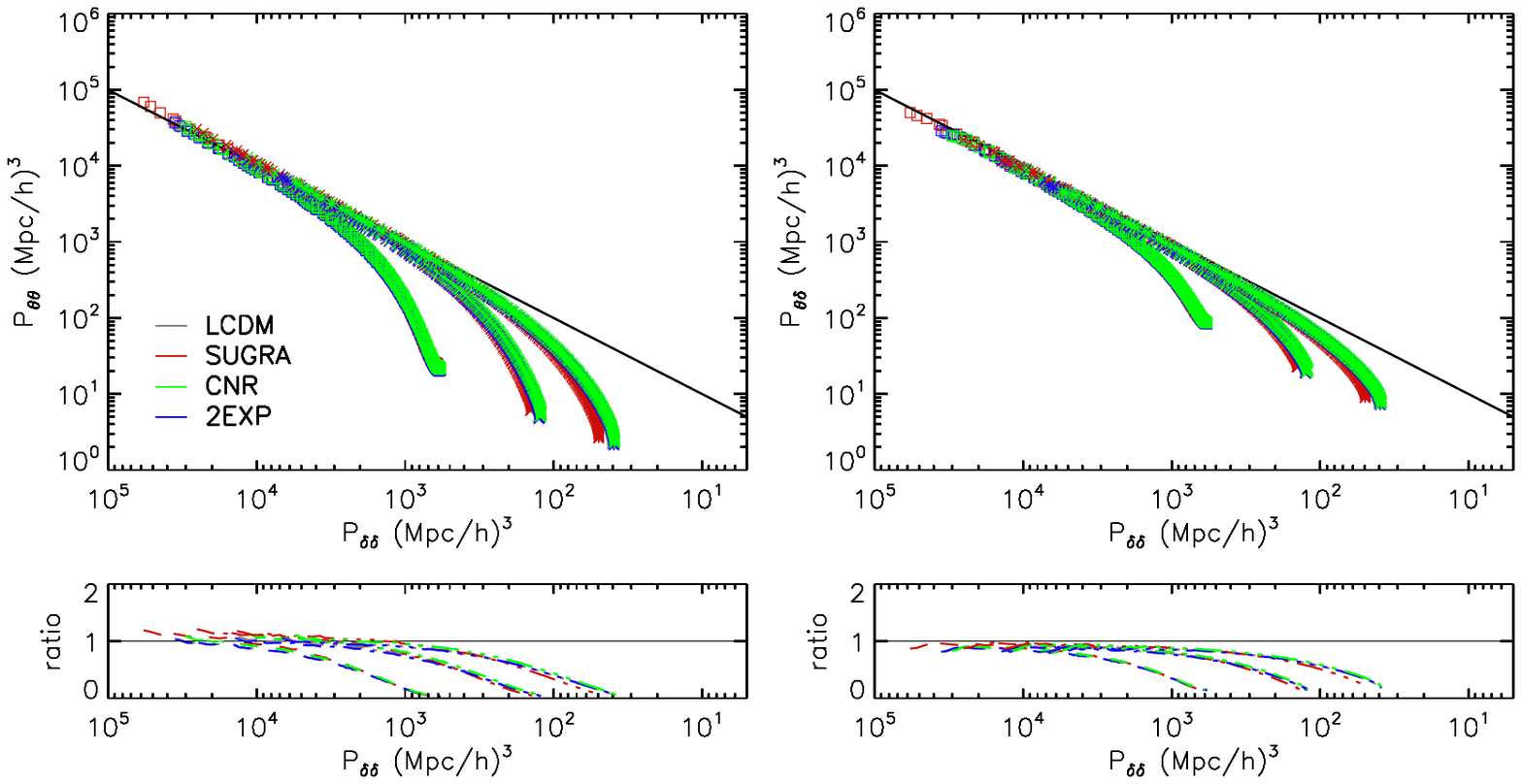}}
 \caption{
Non-linear velocity divergence auto (left) and cross (right) power spectrum plotted as a function of the non-linear matter power spectrum
at $z=0,1$ and 2 in three
quintessence models and $\Lambda$CDM, as labelled. The ratio of the velocity divergence power spectra  to the matter power spectrum at each redshift is plotted in
the smaller panels beneath each main panel.
}\label{allmodels}
\end{figure*}

The left panel in Fig. \ref{ratio} shows the ratio of the power spectra, $P_{\delta \delta}$, $P_{\delta \theta}$ and $P_{\theta \theta}$ 
measured at $z=0$, 
 to the power spectra measured   
at $z=5$ scaled using the ratio of the square of the linear growth factor at $z=5$ and $z=0$ for $\Lambda$CDM. 
It is clear from this plot that all $P(k)$ evolve as expected in linear theory on the largest scales.
Note a linear scale is used on the $x$-axis in this case. In the right panel in Fig. \ref{ratio} all the power spectra have been divided by the
linear theory matter power spectrum measured from the simulation at $z=5$, scaled using the ratio of the linear growth factor at $z=5$ and $z=0$.
This removes the sampling variance from the plotted ratio \citep{1994MNRAS.270..183B}.
In both panels, the error bars represent the scatter over eight simulations in $\Lambda$CDM averaging the 
power spectra after imposing the distortions along  the $x, y$ or $z$ axis in turn. 
 From this figure we can see that the non-linear velocity divergence power spectra can be substantially different from the matter power spectrum on very large scales 
$k\sim 0.03 h$Mpc$^{-1}$. The linear perturbation theory assumption that the velocity divergence power spectra is the same as the matter $P(k)$ is not valid even on
these large scales. In the case of $\Lambda$CDM this difference is $\sim 20\%$ at $k = 0.1 h$Mpc$^{-1}$.
Note in the right panel in Fig. \ref{ratio}, the $10\%$ difference in the ratio of the cross power spectrum to the matter power spectrum, on the largest
scale considered, indicates that we have a biased estimator of $\theta$ which is low by approximately 10\%.

We find that the $P_{\delta \theta}$ and $P_{\theta \theta}$ measured directly from the 
simulation differ from the matter power spectrum by more then was reported by \citet{2009MNRAS.393..297P}.
These authors 
did not measure $P_{\delta \theta}$ and $P_{\theta \theta}$ directly, but instead obtained these quantities by fitting Eq. \ref{10} to the redshift space
monopole power spectrum measured from the simulations. 
In Fig. \ref{ratio2} we plot the same ratios as shown in the right panel of Fig. \ref{ratio} measured from one $\Lambda$CDM (left panel) and SUGRA (right panel) 
simulation. From our simulations it is possible to find a realisation of the density and velocity fields where the measured matter power spectrum and the velocity divergence 
power spectra are similar on large scales.
\begin{figure}
{\epsfxsize=8.truecm
\epsfbox[83 372 435 703]{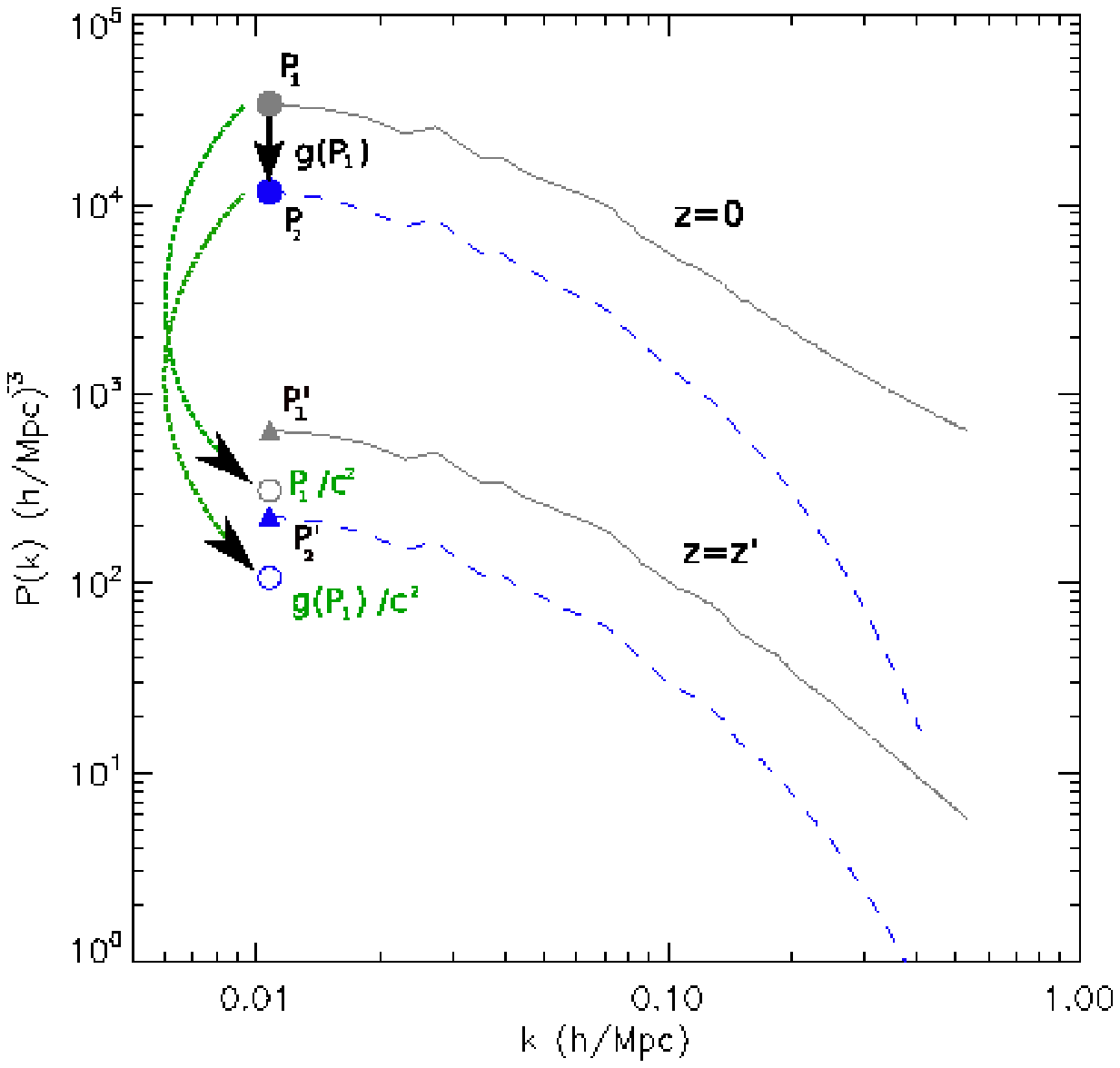}}
 \caption{
A schematic illustration showing how the $z=0$ non-linear matter power  spectrum can be rescaled to find the velocity divergence power spectrum at any redshift
$z=z'$. The upper two curves represent the non-linear matter power spectrum, $P_1$, in grey and the velocity divergence power spectrum, $P_2$, plotted as a blue dashed line,
at $z=0$. The power in the first bin is represented as a filled circle for each spectrum.
The lower two curves, $P'_1$ and $P'_2$,
are the non-linear matter and velocity divergence spectra at $z=z'$. The power in the first bin is represented as a filled triangle in each case.
The fitting formula for $g(P_1)$ (Eq. \ref{g}) generates the non-linear velocity divergence power spectra at $z=0$.
Using the function given in Eq. \ref{c}, the matter power spectrum $P_1$ and $g(P_1)$ can be rescaled to an earlier redshift. The power in the first bin from the
rescaled $P_1$ and $g(P_1)$ are shown as an empty grey and blue circle respectively.
Note that $P_1$ and $P_2$  have been artifically separated for clarity.
}\label{cartoon}
\end{figure}


Having found that the measured $P_{\delta \theta}$ and $P_{\theta \theta}$ differ significantly from $P_{\delta \delta}$, we now test if the grid assignment 
scheme has any impact on our results. As explained in Section \ref{4.2}, the velocity $P(k)$ are computed by taking the Fourier transform
of the momentum field divided by the density field to reduce the impact of the grid assignment scheme \citep{Scoccimarro:2004tg}.
\citet{2009PhRvD..80d3504P} showed that the CIC assignment scheme affects the measured $P(k)$ beyond $\sim$20\% of the Nyquist frequency.
In Fig. \ref{ratio128} we show the power spectrum measurements for four different FFT dimensions to show the scales at which we get a robust measurement.
For $N_{\tiny\mbox{FFT}} = 350$ the power spectra have converged on scales up to $k \sim 0.2 h$Mpc$^{-1}$.


In the top row of Fig. \ref{sugracnr}, the ratios $P^s_0(k)/P^r(k)$ and $P^s_2(k)/P^s_0(k)$ are  plotted as solid lines in the left and right hand panels respectively. In this figure we have
 overplotted as grey dashed lines, the ratio of the redshift space monopole  moment to the real space power spectrum where
\begin{eqnarray}
\label{10}
P^s_0(k) = P_{\delta \delta}(k) + \frac{2}{3}f P_{\delta \theta}(k) + \frac{1}{5}f^2P_{\theta \theta}(k) \, .
\end{eqnarray}
On scales $ 0.05 < k (h$Mpc$^{-1}$$)<0.2$, this model for the redshift space power spectrum reproduces the measured $P_s(k,\mu)$ and
 is a significant improvement compared to Eq. \ref{mr}. This form does not include any
modelling of the damping due to velocity dispersion. The extended model proposed by \citet{Scoccimarro:2004tg} given in Eq. \ref{SM}, which does include damping, 
 is also plotted as a black dot-dashed line
 for $\Lambda$CDM in the top row in Fig. \ref{sugracnr}.
The redshift space quadrupole to monopole ratio in the quasi-linear regime, including the velocity divergence power spectra, is
\begin{eqnarray}
\label{dm}
\frac{P^s_2}{P^s_0} = \frac{\frac{4}{3}f P_{\delta \theta} + \frac{4}{7}f^2 P_{\theta \theta}}{P_{\delta \delta} + \frac{2}{3}f P_{\delta \theta} +\frac{1}{5}f^2 P_{\theta \theta}} \, .
\end{eqnarray}
This model does well at reproducing the ratio of the redshift space to real space power spectrum, although it underpredicts the ratio on scales $k<0.02 h$Mpc$^{-1}$.
The corresponding plots for the SUGRA and CNR models are shown in the middle and bottom rows of Fig. \ref{sugracnr}. 
It is clear that including the velocity divergence power spectrum in the model for $P^s_0$
and $P^s_2$, produces a good fit to the measured redshift space power in both quintessence models on scales up to $k \sim 0.2 h$Mpc$^{-1}$.

\section{ Results II: The density velocity relation \label{RT12} }

In Section \ref{dv} we examine the relationship between the non-linear matter and velocity divergence power spectra in different cosmologies. In Section \ref{rt2} we
study the redshift dependence of this relationship and provide a prescription which can be followed to generate predictions for
 the non-linear velocity divergence power spectrum at a given redshift.

\subsection{Dependence on cosmological model \label{dv}}

The linear continuity equation, Eq. \ref{con}, gives a one to one correspondence between the velocity and density fields with a cosmology dependent factor, $f(\Omega_m,\gamma)$.
Once the overdensities become non-linear, this relationship no longer holds. \citet{1992ApJ...390L..61B} derived the non-linear relation between $\delta$ and 
$\theta$ in the case of an initially Gaussian field.
 \citet{Chodorowski:1996bu} extended this relation
 into the weakly non-linear regime up to third order in perturbation theory and found the 
result to be a third order polynomial in $\theta$. More recently, \citet{2008MNRAS.391.1796B}  found a relation between $\theta$ and $\delta$ using the spherical 
collapse model. In all of these relations, the dependence on cosmological parameters was found to be extremely weak \citep{1992ApJ...390L..61B, Bouchet:1994xp}. 
The velocity divergence depends on $\Omega_m$ 
and $\Omega_{\Lambda}$, in a standard $\Lambda$CDM cosmology, only through the linear growth rate, $f$ \citep{Scoccimarro:1999ed}.

We showed in the previous section that including the velocity divergence auto and cross power spectrum 
accurately reproduces the redshift space power spectrum for a range of dark energy models on 
scales  where the Kaiser formula fails. 
The quantities in Eqs. \ref{dm} and \ref{SM} can be calculated
if we exploit the relationship between the velocity and density field. 
 In  Fig. \ref{allmodels} we plot the velocity divergence 
auto (left panel) and cross (right panel) power spectrum as a function of the matter power spectrum for $\Lambda$CDM and the three quintessence dark energy models. We find that the density 
velocity relationship is very similar for each model at the redshifts considered, with only a slight difference for the SUGRA model at high redshifts and at small scales. 
 The departure of the SUGRA model from the general density velocity relation is due to shot noise, which affects the  power spectrum  most 
at these scales in the SUGRA model 
as it has the lowest amplitude. We have verified that this effect is due to shot noise by sampling half the particles in  the same volume, thereby
 doubling the shot noise, and repeating the $P(k)$ measurement to find an even larger departure.
Fig. \ref{allmodels} shows the independence of the density velocity relation not only of the values of  cosmological parameters, as found in previous works, 
\citet{1992ApJ...390L..61B}, but also a lack of
dependence on the cosmological expansion history and initial power spectrum.

Fitting over the range $0.01<k (h/$Mpc$<0.3)$, we find the following
 function accurately describes the relation between the non-linear velocity divergence and matter power spectrum at $z=0$ to better 
than $5\%$ on scales $k<0.3 h$Mpc$^{-1}$, 
\begin{eqnarray}
\label{g}
P_{x y}(k) = g(P_{\delta \delta}(k)) = \frac{\alpha_0\sqrt{P_{\delta \delta}(k)}  +\alpha_1 P_{\delta \delta}^2(k)}{\alpha_2 + \alpha_3 P_{\delta \delta}(k)} \, ,
\end{eqnarray}
where $P_{\delta \delta}$ is the non-linear matter power spectrum. For the cross power spectrum 
$P_{x y} = P_{\delta \theta}$,  
$\alpha_0 = -12288.7$, $\alpha_1 = 1.43$, $\alpha_2 = 1367.7$ and $\alpha_3 = 1.54$ 
and for $P_{x y} = P_{\theta \theta}$,  
$\alpha_0 = -12462.1$, $\alpha_1 = 0.839$, $\alpha_2 = 1446.6$ and $\alpha_3 = 0.806$; 
all points were weighted equally in the fit and the units for  $\alpha_0, \alpha_1$ and $\alpha_3$ are (Mpc$/h)^{3/2}$, (Mpc$/h)^{-3}$ and (Mpc$/h)^{-3}$ respectively.
The power spectra used for this fit are the average $P_{\theta \theta}$, $P_{\delta \theta}$ and $P_{\delta \delta}$ measured from 
eight $\Lambda$CDM simulations.

\subsection{Approximate formula for $P_{\delta \theta}$ and $P_{\theta \theta}$ for arbitrary redshift \label{rt2}} 

In perturbation theory, the solution for the density contrast is expanded as a series around the background value.
\citet{Scoccimarro:1997st} found the following solutions for $\delta$ and $\theta$ to arbitrary order in perturbation theory,
\begin{eqnarray}
\label{sep}
\delta(k,\tau) = \sum_{n=1}^{\infty} D_n(\tau) \delta_n(k) \nonumber
\\ 
\theta(k,\tau) = \sum_{n=1}^{\infty} E_n(\tau) \theta_n(k) \, ,
\end{eqnarray}
where $\delta_1(k)$ and $\theta_1(k)$ are linear in the initial density field, $\delta_2$ and $\theta_2$ are quadratic in the initial density field etc.
\citet{Scoccimarro:1997st} showed that using a simple approximation to the
equations of motion, $f(\Omega_{\rm m}) = \Omega_{\rm m}^{1/2}$, the equations become separable and  $E_n(\tau) = D_n(\tau) = D(\tau)^n$, where 
$D(\tau)$ is the linear growth factor of density perturbations.
We shall use these solutions for $\delta(k,\tau)$ and $\theta(k,\tau)$
to approximate the redshift dependence of the density velocity relation found in Section \ref{dv}. This relation
does not depend on the cosmological model but we shall assume a $\Lambda$CDM cosmology and find the approximate redshift dependence as
a function of the $\Lambda$CDM linear growth factor.

The fitting function given in Eq. \ref{g} generates the non-linear velocity divergence power spectrum, $P_{\delta \theta}$ or  $P_{\theta \theta}$ from the non-linear matter power spectrum, 
$P_{\delta \delta}$ at $z=0$. Fig. \ref{cartoon} shows a simple illustration of how the function $g(P_{\delta \delta})$ and $P_{\delta \delta}$ at $z=0$ can be rescaled to give
the velocity divergence power spectra at a higher redshift, $z'$. Using the simplified notation in the diagram, where $P_1 = P_{\delta \delta}$, and given the function 
$g(P_{\delta \delta})$,
we can find a redshift dependent function, $c(z)$, with which to rescale $g(P_{\delta \delta}(z=0))$ to the velocity divergence $P(k)$ at $z'$. 
At the higher redshift, $z'$, the non-linear 
matter and velocity divergence power spectra
 are denoted as 
$P'_1$ and $P'_2$ respectively in Fig. \ref{cartoon}.


Using the  solutions in Eq. \ref{sep}, to third order in perturbation theory, see Appendix \ref{A}, we assume a simple expansion 
with respect to the initial density field, 
to find the following ansatz for the mapping $P'_1(z=z') \to P'_2(z=z')$ which can be approximated as $ P_1(z=0)/c^2(z=0,z') \to g(P_1)/c^2(z=0,z')$
where
\begin{eqnarray}
\label{c}
c(z, z') = \frac{D(z) + D^2(z) +D^3(z)}{D(z') + D^2(z')+ D^3(z')} \, ,
\end{eqnarray}
and $D(z)$ is the linear growth factor.
The equivalence of these mappings gives $P'_1 -P'_2 =
(P_1-g(P_1))/c^2$ which allows us to calculate $P'_2$ at $z=z'$ if we have $P_1(z=0)$, $g(P_1(z=0)$ and $P'_1(z=z')$.
Writing this  now in terms of $P_{\delta \delta}$, instead of $P_1$, we have the following equation
\begin{eqnarray}
\label{fullmodel}
P_{x y }(k,z') = \frac{ g(P_{\delta \delta}(k,z=0))  - P_{\delta \delta}(k,z=0)}{ c^2(z=0, z') } \nonumber \\
+ P_{\delta \delta}(k,z') \, ,
\end{eqnarray}
where $g(P_{\delta \delta})$ is the function in Eq. \ref{g} 
 and $P_{x y }$ is either the nonlinear cross or auto power spectrum, 
$P_{\delta \theta}$ or $P_{\delta \delta}$.

\begin{figure*}
{\epsfxsize=17.truecm
\epsfbox[79 442 546 685]{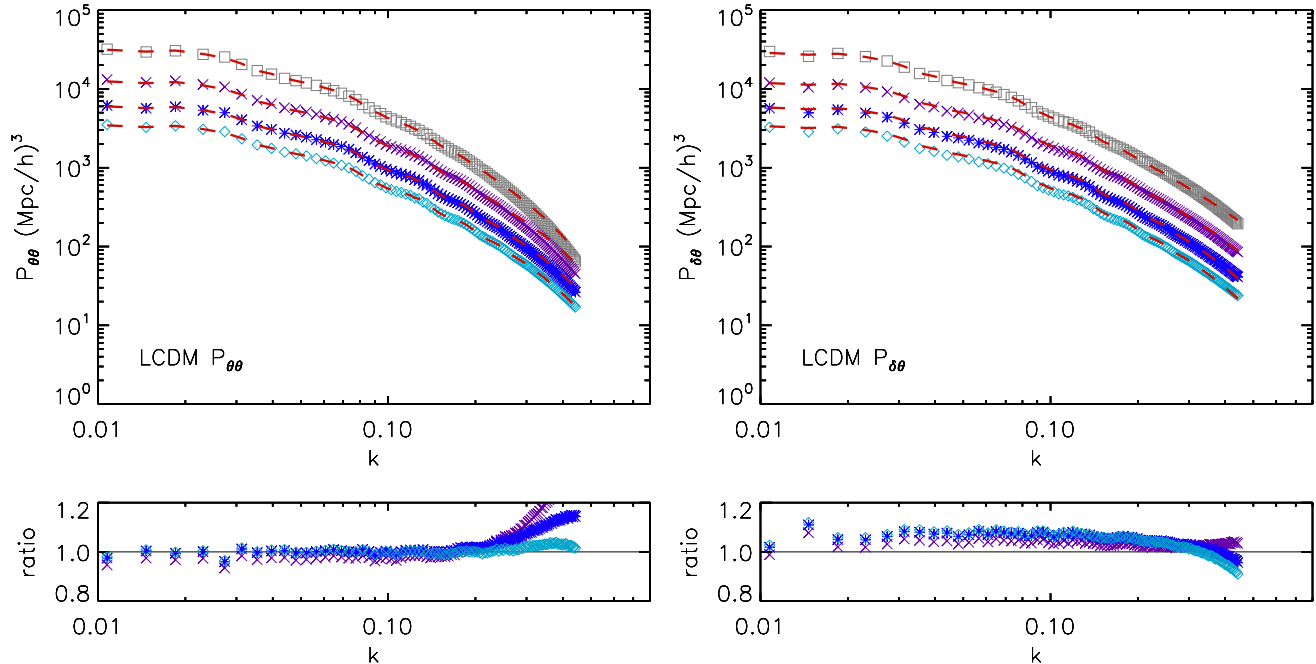}}
 \caption{
Non-linear velocity divergence auto and cross power spectrum,
in the left and right panels respectively, measured from the  $\Lambda$CDM simulations
at $z=0$ (open grey squares),  $z=1$ (purple crosses),  $z=2$ (blue stars) and  $z=3$ (cyan diamonds).
Overplotted as red dashed lines is the function given in
Eq. \ref{fullmodel} at redshifts $z=1,2$ and 3. The lower panels show the function in Eq. \ref{fullmodel}
divided by the measured spectra at $z=1,2$ and 3.
}\label{pscaling}
\end{figure*}

In the left panel of Fig. \ref{pscaling}, we plot the $\Lambda$CDM non-linear power spectrum $P_{\theta \theta}$  at $z = 0,1,2$ and 3.  The function given in 
Eq. \ref{fullmodel} is also plotted as
 red dashed lines 
using the factor $c(z,z')$ given in Eq. \ref{c} and the $\Lambda$CDM linear growth factor at redshift $z=0, 1, 2$ and 3 respectively.
The ratio plot shows the difference between the exact $P_{\theta \theta}$ power spectrum and the function given in Eq. \ref{fullmodel}. 
The right panel in Fig. \ref{pscaling} shows a similar plot for the $P_{\delta \theta}$ power spectrum.
In both cases we find very good agreement between the scaled fitting formula and the measured power spectrum.
Scaling the $z=0$ power spectra using this approximation in Eq. \ref{c} reproduces the non-linear 
$z=1,2$ and 3, $P_{\delta \theta}$ to $\sim 5\%$ and  $P_{\theta \theta}$ to better than $ 5\%$ on scales $0.05 < k (h$Mpc$^{-1}$$)<0.2$.
It is remarkable that scaling the $z=0$ fitting formula using $c$ in Eq. \ref{c} works so well at the different redshifts up to $k<0.3 h/$Mpc and is
completely independent of scale.  

To summarise the results of this section we have found that the quadrupole to monopole ratio given in Eq. \ref{dm} and the model in Eq. \ref{SM}, which 
includes the non-linear matter and velocity divergence power spectra at a given redshift $z'$, 
can be simplified by using the following prescription. Assuming a cosmology with a
given linear theory matter power spectrum we can compute the non-linear matter $P(k)$ at z=0 and at the required redshift, $z'$, using,
for example,  
the phenomenological model HALOFIT \citep{Smith:2002dz} or the method proposed by \citet{2009JCAP...03..014C} in the case of quintessence dark energy. 
These power spectra can then be used in Eq. \ref{fullmodel} together
with the function $g$, given in Eq. \ref{g}, and the linear theory growth factor between redshift $z=0$ and $z=z'$ to find the velocity 
divergence auto or cross power spectrum.
As can be seen from Fig. \ref{pscaling} the function given in Eq. \ref{fullmodel} agrees with the measured non-linear velocity divergence 
power spectrum to $\sim 10\%$ for $k<0.3h$Mpc$^{-1}$ and to $<5\%$ for $k<0.2h$Mpc$^{-1}$ for $\Lambda$CDM. We have verifed that this prescription also reproduces $P_{\delta \theta}$ and 
$P_{\theta \theta}$ to an accuracy of
$10\%$ for  $k<0.3h$Mpc$^{-1}$ for the CNR, SUGRA and 2EXP models 
using the corresponding matter power spectrum and linear growth factor for each model.
This procedure simplifies the redshift space power spectrum in
Eq. \ref{SM} and the quadrupole to monopole ratio given in Eq. \ref{dm}. 
For the dark energy models considered in this paper, this ratio provides an improved fit to the redshift space $P(k,\mu)$ compared to the Kaiser formula
 and incorporating the 
density velocity relation eliminates any 
new parameters which need to be measured separately and may depend on the cosmological model.

\section{Conclusions and Summary \label{5.1}}

One of the primary goals of future galaxy redshift surveys
is to determine the physics behind the accelerating expansion of the Universe by making an accurate measurement of the growth rate, $f$, of large scale structure 
\citep{2009ExA....23...39C}. 
Measuring the growth rate with an error of less than $10\%$ is one of the main science goals of Euclid, as
this will allow us to distinguish modified gravity from dark energy models. With an independent measurement of the 
expansion history, the predicted growth rate for a dark energy model  would agree 
with the observed value of $f$ if general relativity holds.

We use simulations of three quintessence dark energy models which have different expansion histories, linear growth rates and power spectra compared to
$\Lambda$CDM. In a previous paper, \citet{2010MNRAS.401.2181J}, we carried out the first fully 
consistent N-body simulations of quintessence 
dark energy, taking into account different expansion histories,  linear theory power spectra and best fitting cosmological parameters $\Omega_{\rm m}$,
$\Omega_{\rm b}$ and $H_0$, for each model. In this paper we examine the redshift space distortions in the SUGRA, CNR and 2EXP quintessence models.
These models are representative of a broader class
of quintessence models which have different growth histories and dark energy densities at early times compared 
to $\Lambda$CDM.
In particular the SUGRA model has a linear growth rate that differs from 
$\Lambda$CDM by $\sim 20\%$ at $z=5$ and the CNR model has high levels of dark energy at early times, $\Omega_{\rm \tiny{DE}} \sim 0.03$ at $z \sim 200$. The 2EXP model has a similar expansion 
history to $\Lambda$CDM at low redshifts, $z<5$, despite having 
a dynamical  equation of state for the dark energy component. For more details on each of the dark energy models see
\citet{2010MNRAS.401.2181J}.

Redshift space distortions observed in galaxy surveys are the result of peculiar velocities 
which are coherent on large scales, leading to a boost in the observed redshift space 
 power spectrum compared to the real space power spectrum \citep{Kaiser:1987qv}. On small scales these peculiar 
velocities are incoherent and give rise to a damping in the ratio of the redshift to real space power spectrum. The Kaiser formula is a prediction of
the boost in this ratio on very large scales, where the growth is assumed to be linear, and can be expressed as a function of
the linear growth rate and bias, neglecting all non-linear contributions. 

In previous work,
using N-body simulations in a periodic cube of $300 h ^{-1}$Mpc on a side, \citet{Cole:1993kh} found that
the measured value of $\beta =f/b$, where $b$ is the linear bias, 
deviates from the  Kaiser formula on wavelengths of $50 h^{-1}$ Mpc or more as a result of these non-linearities.
\citet{Hatton:1997xs} extended this analysis to slightly larger scales using the Zel'dovich approximation combined with a dispersion model where
non-linear velocities are treated as random perturbations to the linear theory velocity. 
These previous studies do not provide an accurate description of the non-linearities in the velocity field for two 
reasons. Firstly,  the Zel'dovich approximation does not model the 
 velocities
 correctly, as it only treats part of the bulk motions. Secondly, in 
a computational box of length $300 h ^{-1}$Mpc, 
the power which determines the bulk flows has not converged. 
In this work we use a large computational box of side $1500 h^{-1}$Mpc, which allows us to measure redshift space distortions on large scales to 
far greater accuracy than in previous work.

In this paper we find that the ratio of the monopole of the redshift space power spectrum to the real 
space power spectrum  agrees with the linear theory Kaiser formula only on extremely large scales $k<0.03 h$Mpc$^{-1}$ in both $\Lambda$CDM 
and the quintessence dark energy models.
We still find significant scatter between choosing different axes as the line of sight, even though 
we have used a much larger simulation box than that employed in previous studies.
As a result we average over the three power spectra, assuming the distortions lie along the $x$, $y$ and $z$ directions in turn,
for the redshift space power spectrum in this paper. Instead of using the measured matter power spectrum in real space, we find that 
the estimator suggested by \citet{Cole:1993kh}, 
involving the ratio of the quadrupole to monopole redshift space power spectrum, works  better than using the monopole and  agrees with the expected linear theory
on slightly smaller scales  $k< 0.07h$Mpc$^{-1}$ at $z=0$ for both $\Lambda$CDM and the quintessence models.

As the measured redshift space distortions only agree with the Kaiser formula on scales $k<0.07h$Mpc$^{-1}$, it is clear that
the linear approximation is not correct on scales which are normally considered to be in the \lq linear regime\rq, $k<0.2 h$Mpc$^{-1}$.
In linear theory, the velocity divergence power spectrum is simply a product of
 the matter power
spectrum and  the square of the linear growth rate. In this work we have demonstrated that non-linear terms in the velocity divergence power spectrum persist on scales 
$0.04< k (h$Mpc$^{-1})<0.2$.  
These results agree with \citet{Scoccimarro:2004tg} who also found significant non-linear corrections
due to the evolution of the velocity fields on large scales, assuming a $\Lambda$CDM cosmology.
We have shown that including the non-linear velocity divergence auto and cross power spectrum in the expression for the redshift space $P(k)$ leads to a 
significant improvement when trying to match the measured quadrupole to monopole ratio for both $\Lambda$CDM and  quintessence 
dark energy models. 

Including the non-linear velocity divergence cross and auto power spectra
 in the expression for the redshift space power spectrum increases the number of parameters needed and depends on the cosmological model 
that is used. Using the non-linear matter and velocity divergence power spectra we have found a density velocity relation which is model independent over a range of redshifts.
Using this relation it is  possible to write the non-linear velocity divergence auto or cross power spectrum at a given redshift, $z'$,
in terms of the non-linear matter power spectrum  and linear growth factor at  $z=0$ and $z=z'$. This formula is given in Eq. \ref{fullmodel} in Section \ref{rt2}.
 We find that this formula  accurately reproduces the non-linear velocity divergence $P(k)$ to within $10\%$ for $k<0.3 h $Mpc$^{-1}$ and to 
better than $5\%$ for $k<0.2 h$Mpc$^{-1}$ for both $\Lambda$CDM and the dark energy models used in this paper. 

It is clear that including the non-linear velocity divergence terms  results in an improved model for redshift space distortions on scales $k<0.2 h$Mpc$^{-1}$ 
for different cosmological models. 
 Current galaxy redshift surveys can provide only very weak constraints on $P_{\delta \theta}$ and $P_{\theta \theta}$ \citep{2002MNRAS.335..887T}.
The relation given in this paper between the non-linear velocity divergence and matter power spectra will be useful for analysing 
redshift space distortions in future galaxy surveys as it removes the need to use noiser and sparser velocity data.

\section*{Acknowledgments} 

EJ acknowledges receipt of a fellowship funded by the European
 Commission's Framework Programme 6, through the Marie Curie Early Stage
Training project MEST-CT-2005-021074.
This work was supported in
part by grants from the Science and Technology Facilities
Council held by the ICC and 
the Institute for Particle Physics Phenomenology at Durham University.
We acknowledge helpful conversations with  Shaun Cole and Martin Crocce.

\bibliographystyle{mn2e}
\bibliography{mybibliography}

\appendix
\section{Approximate formula for $P_{\delta \theta}$ and $P_{\theta \theta}$ for arbitrary redshift \label{A}}

Eq. 18 in this paper relates $P_{xy}(z')-P_{\delta \delta}(z')$ at $z=z'$ to the same expression at redshift $z=0$ using a variable $c^2$.
Note from Eq. 15 $g(P_{\delta \delta}(z = 0)) = P_{xy}(z=0)$ in Eq. 18.
From Eqs. 16 in our paper and using the result by Scoccimarro et al. 1998
we can write the following solutions for $\theta$ and $\delta$ in terms of scalings of the initial density field \citep{Bernardeau:2001qr},
\begin{eqnarray}
\theta(z) &=&  D(z) \theta_1 + D^2(z) \theta_2 + D^3(z) \theta_3 + \cdots
\end{eqnarray}
and
\begin{eqnarray}
\delta(z) &=&  D(z) \delta_1 + D^2(z) \delta_2 + D^3(z) \delta_3 + \cdots \,\,\,.
\end{eqnarray}
\noindent Squaring these expressions and emsemble averaging we can write the velocity divergence power spectrum and the matter power spectrum
to third order in perturbation theory as
\begin{eqnarray}
P_{\theta \theta}(z') &\sim& <|D(z')\theta_1 + D^2(z')\theta_2 + D^3(z')\theta_3|^2>
\\
P_{\delta \delta}(z') &\sim& < |D(z')\delta_1 + D^2(z')\delta_2 + D^3(z')\delta_3|^2 >\,.
\end{eqnarray}
\noindent  Using the fact that $|D\theta_1 + D^2\theta_2 +D^3\theta_3| \leq |D\theta_1| + |D^2\theta_2| + |D^3\theta_3|$  we can approximate this as
\begin{eqnarray}
P_{\theta \theta}(z') \leq <(D(z')|\theta_1| + D^2(z')|\theta_2| + D^3(z')|\theta_3|)^2>
\\
P_{\delta \delta}(z') \leq < (D(z')|\delta_1| + D^2(z')|\delta_2| + D^3(z')|\delta_3|)^2 >\,,
\end{eqnarray}
\noindent and we assume that 
\begin{eqnarray}
&&<|D(z')\theta_1 + D^2(z')\theta_2 + D^3(z')\theta_3|^2> 
\\&& - <(D(z')|\theta_1| + D^2(z')|\theta_2| + D^3(z')|\theta_3|)^2> \sim \nonumber
\\&& < |D(z')\delta_1 + D^2(z')\delta_2 + D^3(z')\delta_3|^2 > \nonumber
\\ &&- < (D(z')|\delta_1| + D^2(z')|\delta_2| + D^3(z')|\delta_3|)^2 > \,. \nonumber
\end{eqnarray}
\noindent Taking the difference of the two power spectra we have
\begin{eqnarray}
&&P_{\theta \theta}(z') - P_{\delta \delta}(z') \sim \nonumber
\\ \nonumber
&&<(D(z')|\theta_1| + D^2(z')|\theta_2| + D^3(z')|\theta_3|)^2> \nonumber
\\
&& - <(D(z')|\delta_1| + D^2(z')|\delta_2| + D^3(z')|\delta_3|)^2>
\end{eqnarray}
and as $x^2-y^2 = (x-y)(x+y)$ we can rewrite this as
\begin{eqnarray}
&&P_{\theta \theta}(z') - P_{\delta \delta}(z') \sim
\\ \nonumber
&& <[D(|\theta_1| -|\delta_1|) + D^2(|\theta_2|-|\delta_2|)+D^3(|\theta_3|-|\delta_3|)]\nonumber
\\ \nonumber
&& \times [ D(|\theta_1| +|\delta_1|) + D^2(|\theta_2| +|\delta_2|)+D^3(|\theta_3| + |\delta_3|)]> \,.
\end{eqnarray}
  Multiplying out the rhs of this equation and denoting the modulus of variable $|x|$ as $x$ for simplicity, we have
\begin{eqnarray}
&& P_{\theta \theta}(z') - P_{\delta \delta}(z') \sim
\\ \nonumber
&&
 <\{
D^2[\theta^2_1 -\delta^2_1] + D^3[(\theta_1 -\delta_1)(\theta_2 +\delta_2) + (\theta_1 +\delta_1)(\theta_2 -\delta_2)]
\\ \nonumber
&& + D^4[(\theta_1 -\delta_1)(\theta_3 +\delta_3) + (\theta^2_2 -\delta^2_2) +(\theta_1 +\delta_1)(\theta_3 -\delta_3)   ]
\\ \nonumber
&&+ D^5[ (\theta_2 -\delta_2)(\theta_3 +\delta_3) + (\theta_2 +\delta_2)(\theta_3 -\delta_3)] +D^6[\theta^2_3 -\delta^2_3]
\}> \, ,
\end{eqnarray}
\noindent and then taking out a factor of $[\theta^2_1 -\delta^2_1]$ on the rhs we have
\begin{eqnarray}
&&P_{\theta \theta}(z') - P_{\delta \delta}(z') \sim
\\ \nonumber
&&< [\theta^2_1 -\delta^2_1]\{ D^2 +D^3[\frac{\theta_2 +\delta_2}{\theta_1 +\delta_1} + \frac{\theta_2 -\delta_2}{\theta_1 -\delta_1}]
\\ \nonumber
 &&+ D^4 [\frac{\theta_3 +\delta_3}{\theta_1 +\delta_1} + \frac{\theta^2_2 -\delta^2_2}{\theta^2_1 -\delta^2_1} +\frac{\theta_3 -\delta_3}{\theta_1 -\delta_1}]
\\ \nonumber
&&+ D^5 [2\frac{\theta_3\theta_2 -\delta_3\delta_2}{\theta^2_1 -\delta^2_1}] + D^6[\frac{\theta^2_3 -\delta^2_3}{\theta^2_1 -\delta^2_1}]
\} >\, .
\end{eqnarray}

\noindent As $\theta_1$ and $\delta_1$ are linear in the initial density contrast, which we assume to be different to the linear density contrast, 
$\theta_1\sim\delta_1 \sim \delta_i$ and  $\theta_2 \sim \delta_2\sim  \delta_i +
\delta^2_i$ is quadratic in the initial density contrast and $\theta_3 \sim \delta_3\sim  \delta_i +
\delta^2_i +\delta^3_i$ is cubic in the initial density field, we
assume $\theta_1 + \theta_2 \sim \delta_1 + \delta_2$, $\theta_1 + \theta_3 \sim \delta_1 + \delta_3$
and $\theta_1 - \theta_2 \sim \delta_1- \delta_2$, $\theta_1 - \theta_3 \sim \delta_1 - \delta_3$
so the fractions in the above equation are unity and
\begin{eqnarray}
P_{\theta \theta}(z') - P_{\delta \delta}(z') &
\\ \sim &<[\theta^2_1 -\delta^2_1]>\{ D^2 + 2D^3+3D^4+2D^5+D^6\}\nonumber
\\ \sim &<[\theta^2_1 -\delta^2_1]>\{D(z')+D^2(z')+D^3(z')\}^2\nonumber
\end{eqnarray}

\noindent Similarly for $P_{\theta \theta}(z) - P_{\delta \delta}(z)$ we have
\begin{eqnarray}
P_{\theta \theta}(z) - P_{\delta \delta}(z)&
\\ \sim &<[\theta^2_1 -\delta^2_1]>\{D(z)+D^2(z)+D^3(z)\}^2 \nonumber
\end{eqnarray}

\noindent Taking the ratio of the two previous equations, the redshift independent factor $ [\theta^2_1 -\delta^2_1]$ cancels and
we obtain the following ansatz
\begin{eqnarray}
\frac{P_{\theta \theta}(z') - P_{\delta \delta}(z')}{P_{\theta \theta}(z) - P_{\delta \delta}(z) } & \sim & \frac{[D(z')+D(z')^2+D(z')^3]^2}{[D(z)+D(z)^2+D(z)^3]^2}
\end{eqnarray}
which is the expression in Eq. 18 in the paper for $z=0$.
A similar approximation works for the cross power spectrum $P_{\delta \theta}$.

\bsp

\label{lastpage}

\end{document}